# A Simulation Algorithm for Brownian Dynamics on Complex Curved Surfaces

Yuguang Yang,[a,b]  Bo Li[a*]

*[a] Institute of Biomechanics and Medical Engineering, Applied Mechanics Laboratory, Department of Engineering Mechanics, Tsinghua University, Beijing 100084, China*

*b Chemical & Biomolecular Engineering, Johns Hopkins University, Baltimore, MD 21218*

**Abstract**

Brownian dynamics of colloidal particles on complex surfaces has found important applications in diverse physical, chemical and biological processes. However, current Brownian dynamics simulation algorithms mostly work for relatively simple surfaces that can be analytically parameterized. In this work, we develop an algorithm to enable Brownian dynamics simulation on extremely complex surfaces. We approximate complex surfaces with triangle mesh surfaces and employ a novel scheme to perform particle simulation on these triangle mesh surfaces. Our algorithm computes forces and velocities of particles in global coordinates but updates their positions in local coordinates, which benefits from the advantages of simulation schemes in both global and local coordinate alone. We benchmark the proposed algorithm with theory and then simulate Brownian dynamics of both single and multiple particles on torus and knot surfaces. The results show that our method captures well diffusion, transport, and crystallization of colloidal particles on complex surfaces with non-trivial topology. This study offers an efficient strategy for elucidating the impact of curvature, geometry, and topology on particle dynamics and microstructure formation in complex environments.

---

[*] Corresponding author.

 E-mail address: libome@tsinghua.edu.cn (B. Li).

# Introduction

Dynamics of micro- and nano-scale particles, cells, and proteins on curved surfaces plays a critical role in a broad range of physical, chemical and biological processes. Common examples include colloidal particle assembly, packing, defect formation, and crystallization on non-flat interface[1-6], protein diffusion on curved membranes[7, 8], cell dynamics (e.g. differentiation and migration) on curved substrates[9-12], and collective locomotion of active particles system on curved surfaces[13-16]. Theoretical approaches including particle simulations[17, 18] and continuum descriptions[19] are constantly employed to understand these dynamical processes. Despite extensive experimental evidence of emerging novel complexity arising from curvature and topology of surface[4, 6, 13, 16, 20], algorithms of particle simulation are largely limited to simple surfaces that are analytically tractable (e.g., spherical, cylindrical, and ellipsoidal surfaces, and other simple geometric primitives)[1, 6, 18, 21]. Establishing a particle-scale simulation algorithm able to tackle surfaces with arbitrary complex geometry can significantly facilitate the understanding of how curvature and topology regulate individual or collective dynamics.

Brownian dynamics (BD) simulation, developed by Ermak and McCammon,[22] has been widely used for micro-sized particle simulation in the Cartesian space. BD simulation captures the hydrodynamics, Brownian forces, and particle interactions and has become an indispensable tool to study the dynamical aspects of colloidal suspensions. Through decades of development, BD simulation has been harnessed to simulate dynamical processes beyond colloids, including cell migration[23], tumor growth[24], protein-protein interaction[25], protein folding[26], and dynamics of polymers [27] and nonspecific particles[28]. Considerable effort has been directed towards extending Brownian simulation or other molecular simulation methods to curved surfaces. Existing approaches can be classified into two categories: a global coordinate scheme and a local coordinate scheme.

As particles moving on a surface can be viewed as imposing a constraint on the positions, the global scheme can be derived from Lagrange classic mechanics[29, 30] formulation under constraints. In such scheme, calculation of forces and positions of particles are carried out in the global coordinates, and the net effect of the constraints is equivalent to adding constraint forces perpendicular to the constraint surface to offset other forces driving particles out of surfaces. The constraint force can either be exactly calculated from Lagrange multipliers or approximated by applying a restoring harmonic potential to confine the simulated particles on the manifold. The resulting algorithms, commonly referred to as constrained BD, allow convenient calculations and analysis via global coordinates and reuse various computational techniques (e.g. parallel computation via domain decomposition) developed for simulations in Euclidean space. However, there are several limitations in this approach. First, the calculation of constraint forces requires smooth surfaces analytically parameterizable, which are generally not available for many complex surfaces. Second, for highly curved surfaces, the update time step should be sufficiently small to ensure constrained forces to stably maintain particles on the surface and have zero tangent component that would otherwise produce incorrect dynamics. The local scheme approach[18, 21] directly evolves the particle dynamics in the local coordinates systems defined by the constraints. Using local coordinates system can guarantee that the constraints are always satisfied without imposing the constraints forces, thus reducing the complications in computing constrained forces and choice of time-step size update. Although the local scheme is generally more accurate and theoretically sound and is appropriate for the study of single particle dynamics on simple analytical curved surface (.e.g. spherical and cylindrical surfaces)[21], the local scheme will induce significant difficulties in complex surface environments, because their local coordinate system parametrizations are usually not available in these situation.

In this paper, we aim to develop a BD simulation algorithm that enables particle-scale simulation on curved surfaces with complex geometry. We approximate a curved surface by triangle mesh surfaces and formulate the simulation algorithm with respect to them. Triangle mesh surfaces [Fig. 1(a) and (b)] have been widely used to approximate complex surface (e.g., organs, bone, and other structures) in computer graphics, mechanics, medicine, and numerical methods like finite element method[19, 31]. The methodology to create and process triangle mesh surfaces are well-established, including reconstructing complex surfaces from real-world 3D experiments[32]. Because their accuracy can be arbitrarily increased by subdividing triangles[33], such triangle mesh surfaces can be used as an approximation to simulate Brownian motion on a wealth of smooth surfaces, ranging from flat surfaces to spherical surfaces, and to complex surfaces with non-trivial topology.

We implement a hybrid manner that involves elements in both the global and local scheme mentioned above. As we will show, the use of a triangle mesh surface enables us to easily keep track of both global coordinates and local coordinates of all particles. Therefore, our algorithm chooses to calculate force and associated velocities using global coordinates, which is, in general, more convenient than using local coordinates. We then choose to update particles' positions in the local coordinates, which is more convenient and avoids computing constraint forces and related complications mentioned above (i.e., small time step and non-zero tangent component).

This paper is organized as follows. We first review the basic geometry and coordinates transformations in triangle mesh surfaces. We then adapt the constrained BD to triangle mesh surface to compute local velocities from forces in the global coordinates and introduce a 'velocity folding' scheme to update positions in the local coordinates. To verify our algorithm, we benchmark the algorithm with theory by simulating Brownian motion on a flat surface and a spherical surface. Subsequently, we demonstrate our algorithm by simulating single and multiple particle dynamics on surfaces with non-trivial topology. The effect of curvature and topology on BD are illustrated. Finally, we conclude the present paper and provide a discussion on possible extensions based on our algorithm.

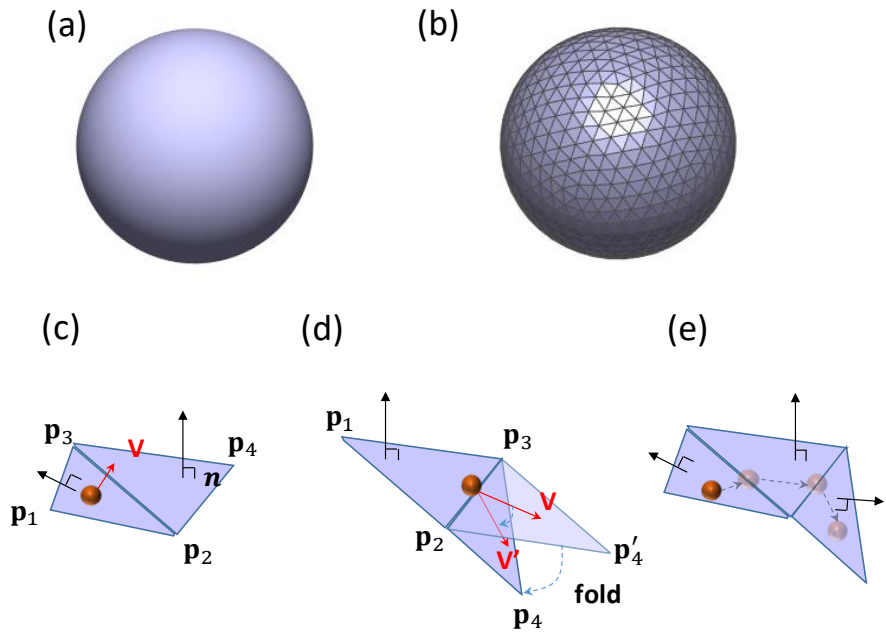

**Figure 1.** (a) A smooth spherical surface. (b) Approximation of a spherical surface by triangle mesh surface. (c) The local coordinate system determined by the vertices of the triangle $p_1$, $p_2$, and $p_3$. A particle lying on the triangle surface with tangent velocity vector **v**. (d) Illustration of velocity folding. Consider a particle lying on the edge $p_2p_3$. We can view this particle is lying on the surface of ($p_1$, $p_2$, $p_3$) with tangent velocity **v**, and simultaneously lying on the surface ($p_2$, $p_3$, $p_4$) with tangent velocity **v'**. The tangent velocity **v'** will be used to evolve particle position on the surface ($p_2$, $p_3$, $p_4$). The tangent speed **v'** can be obtained by fixing the starting point of **v** and then fold the face ($p_2$, $p_3$, $p_4'$) to ($p_2$, $p_3$, $p_4$). (e) An example trajectory of a particle moving on the triangle meshes with its velocity folded twice.

## Theory and Algorithm

### Coordinate systems on a triangle mesh

To facilitate our subsequent discussion, we start with a brief introduction to the coordinate systems on a triangle mesh surface. A triangle mesh surface [Fig. 1(b)] can be described by a vertex set consisting of vertices $p_1$, $p_2$,..., $\in R^3$, and a face set with each triangle face determined by three vertices ($p_i$, $p_j$, $p_k$). The position of a particle lying on the face of the triangle surface can be described by both a global/lab coordinate $\mathbf{r} = (r_1, r_2, r_3)^T$, and a local coordinate $\mathbf{q} = (q_1, q_2)^T$ determined by the three vertices ($\mathbf{p}_1$, $\mathbf{p}_2$, $\mathbf{p}_3$) of the triangle face [Fig. 1(c)]. The local coordinate of every point in the interior of a triangle ($\mathbf{p}_1$, $\mathbf{p}_2$, $\mathbf{p}_3$) is the barycentric coordinate given by

$$\mathbf{r} = (\mathbf{p}_2 - \mathbf{p}_1, \ \mathbf{p}_3 - \mathbf{p}_1) \cdot \mathbf{q} + \mathbf{p}_1, \tag{1}$$

with the constraints $q_1 \geq 0$, $q_2 \geq 0$, $q_1 + q_2 \leq 1$. From Eq. (1), the straightforward coordinate transformation from local $\mathbf{q}$ to global $\mathbf{r}$ is described by

$$\mathbf{r} = \mathbf{J} \cdot \mathbf{q} + \mathbf{p}_1, \mathbf{J} = [\mathbf{p}_2 - \mathbf{p}_1, \mathbf{p}_3 - \mathbf{p}_1], \tag{2}$$

where $\mathbf{J} \in \mathbf{R}^{3 \times 2}$ is usually referred to as Jacobian matrix. Similarly, the transformation from the global $\mathbf{r}$ to local $\mathbf{q}$ is given as

$$\mathbf{q} = \mathbf{J}^* \cdot (\mathbf{r} - \mathbf{p}_1), \ \mathbf{J}^* = (\mathbf{J}^T \cdot \mathbf{J})^{-1} \cdot \mathbf{J}^T, \tag{3}$$

where $\mathbf{J}^* \in \mathbf{R}^{3 \times 2}$ is the pseudo inverse of the original Jacobian matrix $\mathbf{J}$.

The tangent velocity associated with the particle on a surface can also have a local description $\mathbf{v_q}$ and a global description $\mathbf{v_r}$, with the transformation rule given by

$$\mathbf{v_q} = \mathbf{J}^* \cdot \mathbf{v_r}. \tag{4}$$

Eq. (4) can be derived by taking time derivative on the two sides of Eq. (3) (note that $\mathbf{p}_1$ is a constant vector).

## Constrained BD on a triangle mesh

Consider $N$ particles with position vectors $\mathbf{r}_i \in \mathbf{R}^3$ ($i = 1, 2, \ldots, N$) and the constraints on the positions $C(\mathbf{r}_i) = 0$ such that particles are restricted to the surface of interest. The overdamped motion of particles under constraints is governed by[27, 28, 34],

$$\frac{d\mathbf{r}_i}{dt} = \beta \mathbf{D}_i \cdot (\mathbf{F}_i^P + \mathbf{F}_i^B + \mathbf{F}_i^C), \tag{5}$$

where $\beta = (kT)^{-1}$ is the inverse thermal energy, $\mathbf{D}_i$ is a 3×3 diffusivity tensor, $\mathbf{F}_i^P \in \mathbf{R}^3$ denotes unconstrained deterministic forces (e.g., forces arising from particle–particle interaction, particle–wall interaction, particle–field interaction, or active forces), $\mathbf{F}_i^B \in \mathbf{R}^3$ denotes Brownian forces, and $\mathbf{F}_i^C \in \mathbf{R}^3$ denotes constraint forces. The constraint forces are perpendicular to the local tangent plane of the surface at $\mathbf{r}_i$ [34], and they are introduced to ensure the resulting dynamics governed by Eq. (5) is satisfying the constraints $C(\mathbf{r}_i) = 0$. On the triangle mesh surface, $\mathbf{F}_i^C$ is in the direction of $\mathbf{n}_i$, where $\mathbf{n}_i$ is the normal of triangle face in which the particle $i$ lies. Requiring the constraint $C(\mathbf{r}_i) = 0$ to hold all the time gives rise to

$$\frac{\partial C}{\partial t} = \left[\nabla_{\mathbf{r}_i} C\right]^\mathrm{T} \cdot \frac{\mathrm{d}\mathbf{r}_i}{\mathrm{d}t} \propto \mathbf{n}_i^\mathrm{T} \cdot \frac{\mathrm{d}\mathbf{r}_i}{\mathrm{d}t} = 0, \qquad (6)$$

where we have used the fact that $\mathbf{n}_i$ has the same direction as $\nabla_{\mathbf{r}_i} C(\mathbf{r}_i)$ on the position $\mathbf{r}_i$. Plugging Eq. (5) into (6) and requiring $\mathbf{F}_i^C$ to be colinear with $\mathbf{n}_i$, we can derive $\mathbf{F}_i^C$ as

$$\mathbf{D}_i \cdot \mathbf{F}_i^C = -\mathbf{n}_i \cdot (\mathbf{n}_i \cdot \mathbf{n}_i^\mathrm{T})^{-1} \cdot \mathbf{n}_i^\mathrm{T} \cdot \mathbf{D}_i (\mathbf{F}_i^D + \mathbf{F}_i^B). \qquad (7)$$

The equation of motion given by Eq. (5) then reduces to

$$\frac{\mathrm{d}\mathbf{r}_i}{\mathrm{d}t} = \mathbf{P}_i \cdot [\beta \mathbf{D}_i \cdot (\mathbf{F}_i^P + \mathbf{F}_i^B)], \qquad (8)$$

where

$$\mathbf{P}_i = \mathbf{I} - \mathbf{D}_i \cdot \mathbf{n}_i \cdot (\mathbf{n}_i^\mathrm{T} \cdot \mathbf{D}_i \cdot \mathbf{n}_i)^{-1} \cdot \mathbf{n}_i. \qquad (9)$$

When tensor $\mathbf{D}_i$ is diagonal, $\mathbf{P}_i$ reduces to $\mathbf{P}_i = \mathbf{I} - \mathbf{n}_i \cdot \mathbf{n}_i^\mathrm{T}$, which is the familiar orthogonal projector onto the local tangent plane with the normal $\mathbf{n}_i$. Since our triangle mesh surface has zero curvature everywhere except at edges (however, edges have zero measure), we do not need to consider the additional correction of curvature effects of constraints[2].

In constrained BD, the Brownian forces $\mathbf{F}_i^B$ need to have zero value along the direction $\mathbf{n}_i$ [34], which can be obtained by projecting the un-projected Brownian force $\mathbf{F}_i^{B'}$ onto the local tangent plane via

$$\mathbf{F}_i^B = (\mathbf{I} - \mathbf{n}_i \mathbf{n}_i^T) \cdot \mathbf{F}_i^{B'}, \quad (10)$$

where $\mathbf{F}_i^{B'}$ has zero mean and variance given by

$$\langle \mathbf{F}_i^{B'}(t) \mathbf{F}_i^{B'}(t') \rangle = 2\beta^{-2}[\mathbf{D}_i]^{-1} \delta(t-t'), \quad (11)$$

with $\delta$ being the Kronecker delta. It turns out our algorithm has the most intuitive interpretation in the velocity form. Rewrite the constrained dynamics of Eq. (8) as

$$\frac{d\mathbf{r}_i}{dt} = \mathbf{P}_i \cdot \mathbf{v}_i^{uc}, \quad (12)$$

where $\mathbf{v}_i^{uc} = \beta \mathbf{D}_i \cdot (\mathbf{F}_i^P + \mathbf{F}_i^B)$ is the velocity when there are no constraints on particles. Therefore, the effect of constraints is simply the projection of the original velocity onto the local tangent plane via projection operator $\mathbf{P}_i$.

**Algorithm**

Our simulation algorithm has the following key steps during each step for updating particle positions (below we leave out the subscript *i*):

(1) Calculate all unconstrained forces and the resulting velocity $\mathbf{v}^{uc}$ (Eq. (12)) in global coordinate for each particle (note that Brownian forces needs to be generated by Eq. (10)).

(2) Transform $\mathbf{v}^{uc}$ to tangent velocity $\mathbf{v_r}$ via projection $\mathbf{v_r} = \mathbf{P} \cdot \mathbf{v}^{uc}$; then get the local description $\mathbf{v_q}$ via transformation $\mathbf{v_q} = \mathbf{J}^* \cdot \mathbf{v_r}$.

(3) Update local coordinates using local tangent velocity $\mathbf{v_q}$ and Euler-Maruyama scheme[35] via $\mathbf{q}(t + \Delta t) = \mathbf{q}(t) + \mathbf{v_q}\Delta t$.

(4) Transform updated local coordinates $\mathbf{q}(t + \Delta t)$ back to global coordinates r $\mathbf{r}(t + \Delta t)$ via Eq. (2)

One critical procedure in updating local coordinates in step (3) is to handle the case that a particle is moving from one triangle to another triangle, as showed in [Fig. 1(c) and (d)]. When the particle is moving to the edge between two triangles at $\tilde{t}$, we fold the original tangent velocity $\mathbf{v}$ to a new tangent velocity $\mathbf{v}'$ [Fig. 1(d)], which has the same magnitude but different direction, and continues the evolution of $\mathbf{q}$ for the remaining time. We refer to such a method as 'velocity folding' hereafter. The critical rationale behind this velocity folding is the fact that a particle on the surface under no force moves along geodesic according to Lagrange classic mechanics [29, 36]. A geodesic is the shortest path between two points on the surface, which is the generalization of straight lines on a plane to shortest paths on surfaces. As we updating $\mathbf{q}(t + \Delta t) = \mathbf{q}(t) + \mathbf{v_q}\Delta t$, we can view this particle experiences no force and moves with constant velocity. When we use folded velocities as the particle transverses across different faces, we can ensure that a particle moves along a geodesic with constant speed on the triangle mesh surface during the whole interval $\Delta t$. We verify this velocity folding by a simple numerical experiment [Fig. S1]: we initialize a particle with unit tangent velocity on a triangle mesh surface approximating a spherical surface of unit radius; we use a large integration time step of $2\pi$, and after enough folding, the particle moves back to its original position with a negligible error.

The edge hitting time and the neighbor face being hit during the interval $\Delta t$ can be determined by solving

$$t_1 = \frac{q_2}{v_{q,2}}, \ t_2 = \frac{(1-q_1-q_2)}{v_{q,1}+v_{q,2}}, \ t_3 = \frac{q_1}{v_{q,1}}. \tag{13}$$

Among $t_1$, $t_2$, and $t_3$, if $t_1$ is the minimum positive one, the hitting edge is $\mathbf{p}_1\mathbf{p}_2$; if $t_2$ is the minimum positive one, the hitting edge is $\mathbf{p}_2\mathbf{p}_3$; if $t_3$ is the minimum positive one, then the hitting edge is $\mathbf{p}_3\mathbf{p}_1$.

After determining which face will be folded to, we can calculate the after-folding velocity by multiplying the before-folding velocity by a rotation matrix, since the velocity folding from one triangle to another one is equivalent to applying a rotation. The rotation matrix is defined by the rotation angle between the two surfaces and the rotational axis (which is the edge joining the two surfaces).

The following algorithm summarizes the position update procedure on a triangle mesh at each time step.

| Algorithm: BD algorithm on a triangle mesh surface |
| --- |
| **For** each time step |
| Calculate velocities **v** (due to deterministic forces and Brownian forces) in lab coordinate for all particles. |
|     **For** each particle $i$ |
|       Project its velocity in lab coordinate to local velocity description. |
|       Update the local coordinate of particle $i$ with velocity folding. |
|       Transform the particle's local coordinate to global coordinate using Eq. (2) |
|     **End For** |

**End For**

Remarks:

1. All the matrices used for coordinate transformations and velocity folding can be pre-computed and re-used to accelerate the simulation.

2. There are a number of well-developed software to process mesh surfaces (e.g. MeshLab[†], and C++ library libigl[‡]), and analyze dynamics and path on mesh surfaces (MATLAB). Particularly, MeshLab can be used to conveniently scale, deform, rotate, refine, and coarsen the mesh surface.

3. Our C++ implementation of this algorithm is also available[§].

# Methods

In all numerical experiments, we use particle radius $a$ = 1000 nm, diffusivity $D$ = 2.145 $a^2$/s, absolute temperature T= 293 K, and integration time step $\Delta t$=0.0001 s. The characteristic time scale associated with diffusion is $\tau = a^2/D \approx 0.47$ s. All length in this work is measured in the scale of radius $a$. In simulation of multiple particles, we introduce electrostatic repulsive interactions and depletion attraction between particles, which are given by[37],

$$\mathbf{F}_i = -\sum_{j \neq i} \nabla u_{i,j}(r_{ij}),$$
$$u_{i,j}(r_{ij}) = B\exp\left[-\kappa(r_{ij} - 2a)\right] + \Delta\Pi V_{ex}(r_{ij}), \quad (14)$$

where we use electrostatic pre-factor $B$ = 2.29$a/k$T and Debye length $\kappa^{-1}$ = 20 nm, $r_{ij}$ is particle pair separation, and $V_{ex}$ is the excluded volume between spheres[38], which are given by,

$$V_{ex}(r_{ij}) = \frac{4\pi}{3}(a+L)^3 \left[ 1 - \frac{3}{4}\left(\frac{r_{ij}}{a+L}\right) + \frac{1}{16}\left(\frac{r_{ij}}{a+L}\right)^3 \right]. \quad (15)$$

---

[†] http://www.meshlab.net/
[‡] https://libigl.github.io/
[§] https://github.com/yangyutu/BrownianMotionManifold.git

$\Delta\Pi$ is the osmotic pressure difference between the bulk and the excluded volume region. We use $\Delta\Pi = 5.8\times10^{-6}$ $kT$/nm$^3$ and $L = 0.2a$ such that there exist about $5k$T pair attractions between particles. In the simulation, we store each particle's lab coordinate, local coordinate, and the triangle face index. Given simulated trajectories, we construct the position distribution on the surface by counting and normalizing the frequency of the particle appearing inside each face of the mesh surface. To construct displacement distribution on the flat and the spherical surfaces, we initiated 400,000 trajectories at the same initial position and run the simulation up to the observation time.

The basic geometry information on the four triangle mesh faces used in the following sections is listed below.

|  | flat surface | Spherical surface | Torus | Knot |
| --- | --- | --- | --- | --- |
| # of faces | 2288 | 81920 | 56,064 | 449,832 |
| # of vertices | 6864 | 245970 | 168,192 | 224,916 |
| Average face area ($a^2$) | ~0.0014 | ~0.00015 | ~0.018 | ~0.0084 |
| Face length scale($a$) | ~0.05 | ~0.018 | ~0.19 | ~0.13 |
| Total face area ($a^2$) | ~6.24 | ~25 | ~1971 | ~7580 |

Our algorithm can be run on using a single normal desktop CPU core for medium system size (< 1000 particles) within reasonable time. For example, the time for simulating 480 particles on the knot surface for $1\times10^6$ steps take about 20 mins.

## Results and Discussion

## Brownian simulation in a flat plane

We first benchmark our simulation algorithm on a two-dimensional (2D) flat plane. The flat plane is represented by a triangle mesh surface, as shown in Fig. 2(a) and (b). The mesh has average face area of ~$0.0014a^2$, corresponding to a length scale of ~$0.05a$. We simulate a single particle that undergoes Brownian motion, and experiences no other external force [Fig. 2(b)]. Fig. 2 (a) and (b) shows a representative trajectory of a particle starting at origin.

To obtain more quantitative verification on the simulation algorithm, we studied the position distribution of a diffusing particle. For a freely diffusing Brownian particle, its probability distribution at time $t$ is given by the following diffusion equation:

$$\frac{\partial p(\mathbf{r},t)}{\partial t} = D\nabla^2 p(\mathbf{r},t) \tag{16}$$

where $p(\mathbf{r}, t)$ is the probability density at time $t$ and position $\mathbf{r}$, $D$ is the diffusivity parameter that takes the same value of diffusivity used in Brownian motion simulation and $\nabla^2$ is the Laplacian operator. When the diffusion is confined to a plane and the initial starting position is at origin, we have theoretical solution given by

$$p(x,y,t) = \frac{1}{4\pi Dt}\exp\left(-\frac{x^2+y^2}{4Dt}\right), \tag{17}$$

which is the 2D Gaussian distribution with variance $4Dt$.

The solution Eq. (17) can also be parametrized by distance $r$ defined as $r = \sqrt{x^2+y^2}$,

$$p(r,t) = \frac{r}{2Dt}\exp\left(-\frac{r^2}{4Dt}\right). \tag{18}$$

Fig. 2(c) and (d) shows the position distribution constructed from 2D histogram of multiple realizations of BD simulations at different observation time $t$. The symmetric shape of the distribution agrees with expectation. We compare the 1D distribution constructed from simulation trajectory with theoretical prediction in Eq. (18), which show excellent agreement between simulation and theory at different observation time.

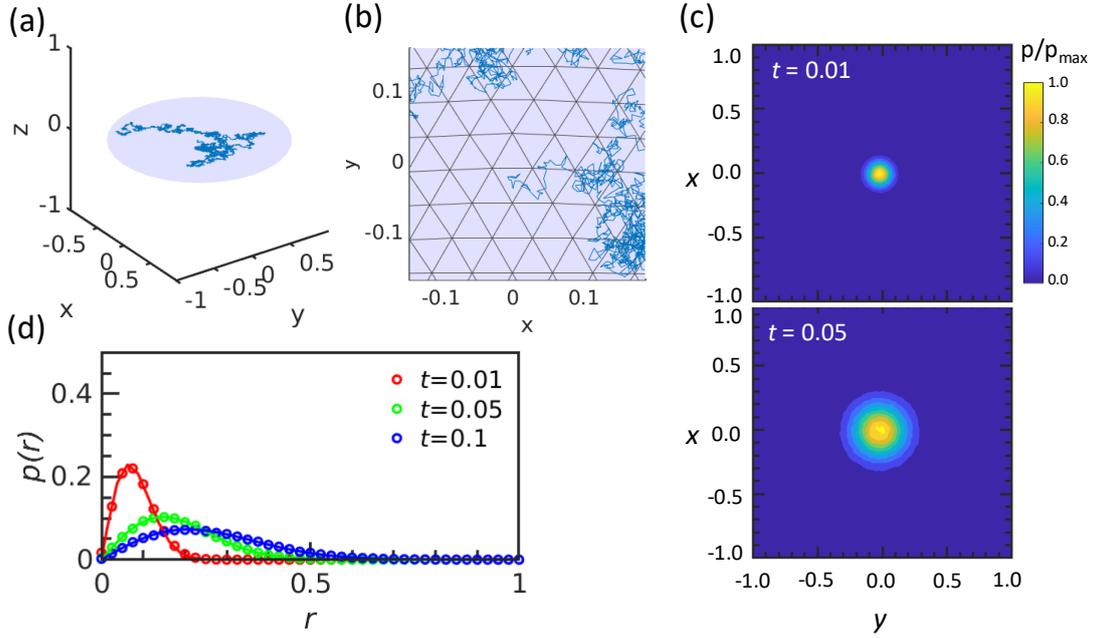

**Figure 2.** (a, b) A representative 1s simulated trajectory of a freely-diffusing colloidal particle on a flat 2D triangle mesh surface embedded in 3D space. The magnified top view is showed in (b). (c) Position distribution on the 2D flat surface at different observation time t=0.01 and 0.05. The distribution is constructed from simulation using trajectories starting at the origin. (d) Position distribution parametrized by distance r constructed from simulation (symbols) and predicted by theory (Eq. (18); solid lines) at different observation time $t$=0.01 s, 0.05 s, and 0.1 s.

## Brownian simulation in spherical surface

We next verify our algorithm by simulating a freely-diffusing particle on a spherical surface of radius $a$. We use a sufficiently refined mesh surface whose triangle faces have a length scale of ~0.018$a$. Fig. 3(a) shows a typical realization of the Brownian simulation of 10 s, with its starting position at the north pole. We then constructed the displacement distribution on the surface from multiple trajectories using histogram method on the mesh (see Methods). The distribution exhibits symmetric shape around the starting position, which agrees well with our expectation [Fig. 3(b)].

To quantify the accuracy of our algorithm, we benchmark the displacement distribution with theory. For a particle undergoing Brownian motion on a spherical surface, the probability density of its position is captured by the same diffusion equation in Eq. (16). If the particle starts at the north-pole, due to its uniform sampling in the azimuth angles $\phi$ coordinate, the evolution of its position density in polar angle $\theta$ coordinate is given by [18]

$$p(\theta,t) = \sum_{l \geq 0}^{\infty} \frac{2l+1}{4\pi R^2} P_l(\cos\theta) \exp(-\frac{l(l+1)Dt}{R^2}), \tag{19}$$

where $P_l$ are the Legendre polynomials with degree $l$ and $R$ is the radius of the sphere. Fig. 3(c) shows the position distribution parameterized by $\cos\theta$, showing good agreement between simulation and theory (Eq. (18)) at different observation time.

We further compare the mean-square angular displacement (MSAD) $\langle \theta(t)^2 \rangle$ of a freely diffusing particle on the sphere, which is defined by

$$\langle \theta(t)^2 \rangle = \left\langle (\cos^{-1}[\frac{\mathbf{r}(\tau+t) \cdot \mathbf{r}(\tau)}{\|\mathbf{r}(\tau+t)\|\|\mathbf{r}(\tau)\|}])^2 \right\rangle_\tau \tag{20}$$

where the expectation is taken by averaging out a long trajectory $\mathbf{r}(t)$.

In the short-time limit where the particle is mainly diffusing on its local neighborhood (e.g., within its own hemisphere) without experiencing the global geometry confinement, the MSAD is characterized by $MSAD = 4Dt/R^2$. In the long-time limit where the particle position $\mathbf{r}(t)$ will be uniformly distributed on the spherical surface, and the MSAD has an asymptote given by[18]

$$\left\langle [\Delta\theta(t \to \infty)]^2 \right\rangle = \frac{1}{4\pi} \int_0^{2\pi} \int_0^{\pi} \theta^2 \sin(\theta) d\theta d\phi = \frac{\pi^2 - 4}{2}. \quad (21)$$

Note that the long-time asymptote is independent of the diffusivity $D$ but depends on the surface geometry (i.e., area). Fig. 3(d) shows the MSAD constructed from a long simulation trajectory, which agrees well with theory in both the short-time and long-time limit. The transition from short-time limit to long-time limit occurs at ~1s, which is also indicated by the MSAD starting to fall below the short-time MSAD curve due to the confinement of surface geometry. This transition time approximately corresponds to a diffusion distance $\sqrt{4Dt} \sim 3a$, where the particle diffuses across the equator towards another pole.

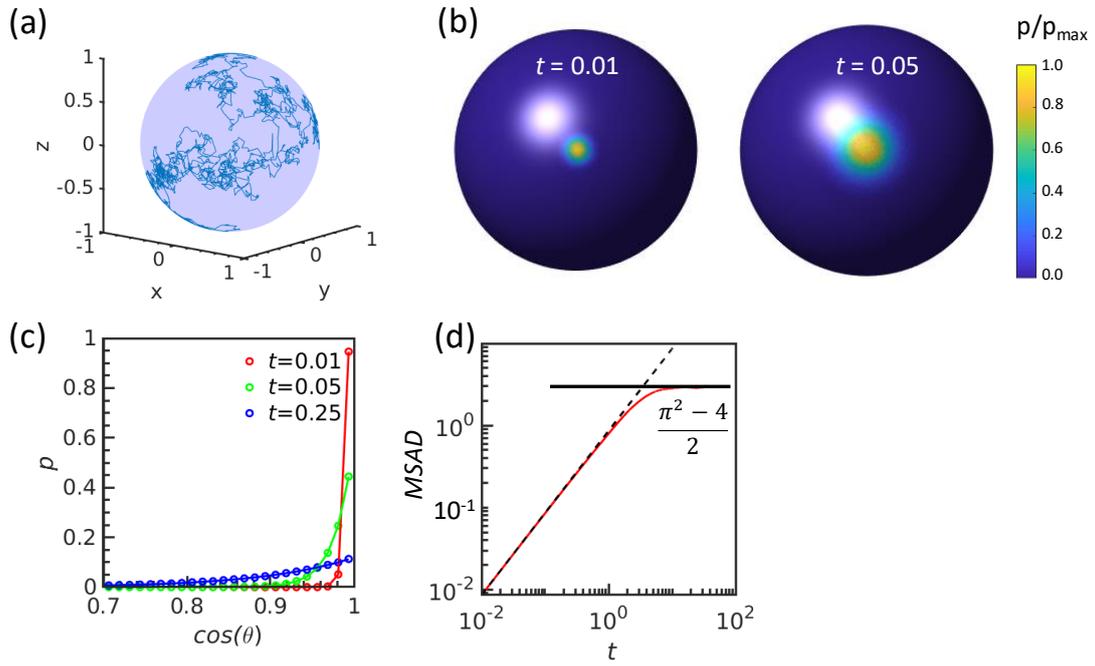

**Figure 3.** (a) A representative 10 s trajectory of a diffusing particle on a spherical surface of radius *a*. (b) Position distribution of a diffusing particle at different observation time *t*=0.01 s (left), 0.05 s (right) on the spherical surface. (c) Position distribution parameterized by $\cos\theta$, where $\theta$ is the polar angle, at different observation time ~~*t*=0.01s, 0.05s,0.25s~~. Symbols denote simulation results and solid lines denote theoretical prediction by Eq.(18). (d) MSAD analysis of a long simulation trajectory (5000 s) on the spherical surface. The dashed line is the theoretical prediction *MSAD* = 4*Dt* that applies to short-time diffusion. The solid horizontal line is the long-time theoretical limit of $(\pi^2 - 4)/2$ (Eq.(21)).

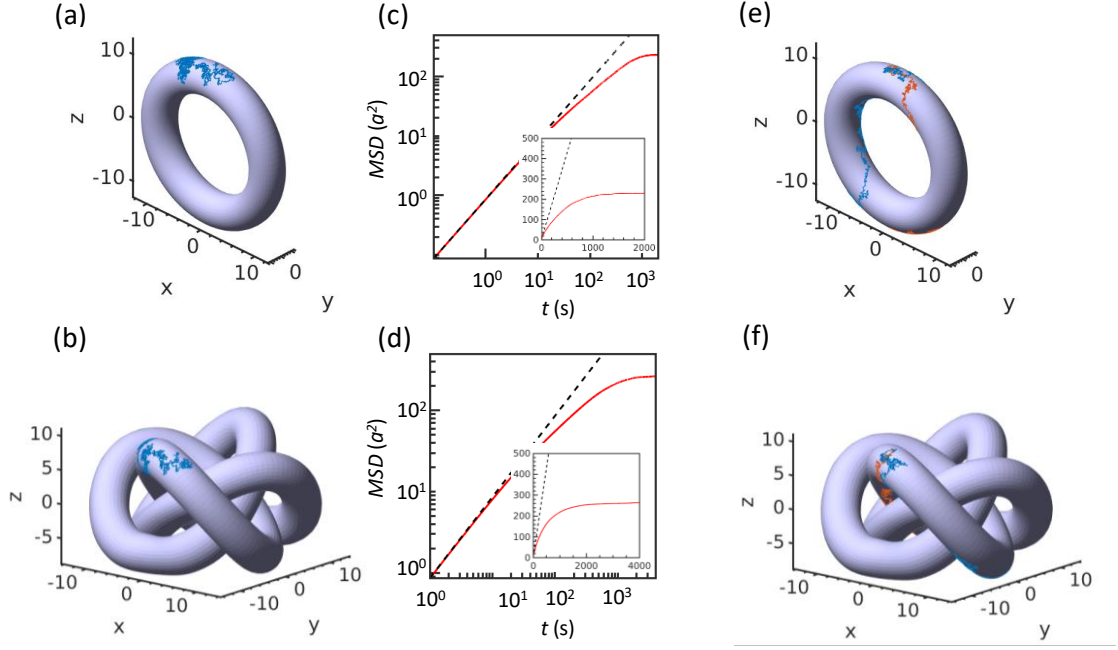

**Figure 4.** Trajectories of 100 s Brownian simulation of single particle diffusing on surfaces with non-trivial topology, including a torus (a) and a knot (b). (c, d) MSD analysis of a long simulation trajectory (10,000 s) on the torus (c) and the knot (d). We calculate the MSD in terms of Euclidean distance (instead of geodesic distance) based on particle lab coordinates. The dashed lines are the theory $MSD = 4Dt$ that applies to diffusion on flat 2D surfaces. Insets are MSD plot in the linear scale. (e,f) Two example trajectories of 100 s Brownian simulation of a single diffusive particle on the torus (e) and a knot (f), where the particle is subject to external field force along negative z direction characterized by $F_z$ = -3 $kT/a$.

## Single particle dynamics on complex surfaces

After benchmarking, we simulate Brownian particles on complex surfaces with non-trivial topology, including a torus and a knot (Fig. 4), to demonstrate the capability of our algorithm. These two surfaces are challenging for previous methods,[17, 18] since they might not have compact parametric forms that allow the computation of constraining forces or direct evolution of dynamics in the local coordinates (the standard torus has analytical form, but its nonstandard variants will not). Fig. 4(a) and

(b) shows the exampled trajectories of simulated single particle freely diffusing on a torus and a knot, respectively. Using the lab coordinates of the particle, the mean squared displacement (MSD) analysis of a long simulation trajectory lasting for 10000s is provided in Fig. 4(c) and (d). On the short time scale $t < \sim 3$ s (corresponding to diffusion distance $\sqrt{4Dt} \sim 5a$ ), the MSD curves on both surfaces can be well approximated by $MSD \approx 4Dt$, since the particle is approximately diffusing on its flat local tangent plane. The impact of curvature emerges on a slightly longer time scale (e.g., $t$ between 10 s and 100 s), where the MSD curves are below the reference curve of $4Dt$ because the distance calculated in 3D lab coordinates is smaller than the traveled distance measured along a curve surface. The impact of global geometry and topology emerges on large time scale $t > 1000$ s, where the $MSD$ plateaus around $\sim 230a^2$ on the torus and, after 1800 s, plateaus around $\sim 270a^2$ on the knot (corresponding to the length scale of $\sim 15a$ and $\sim 16a$, respectively). The MSD takes longer time to plateau at the knot surface (~1800s) than at the torus surface (~1000s) because the knot surface has a larger area (~$7579a^2$) than the torus surface (~$1970a^2$). The square root of the plateau MSD value can either be interpreted as confinement length scale or be interpreted as the average distance between two positions randomly sampled on the surfaces. The topology thus has non-trivial impact on the confinement length scale. Although the knot has a much larger dimension (~ $29a$ by $29a$ by $20a$) than the torus ($25a$ by $5a$ by $25a$), the confinement length scale on the knot is actually comparable to that on the torus. This is because multiple interconnected tube surfaces in the knot enable diffusing particles to have higher chances to get closer to their previously sampled position in terms of distance measured in lab coordinates.

By applying external field force along the negative z direction (e.g., gravity, electric field[39, 40]), our algorithm can be used to study the single particle diffusive transport driven by the interplay of external force, geometry, and topology [Fig. 4(e) and (f)]. Compared with the free diffusion case, particles under external forces quickly move down the surface and equilibrate at the surface bottom. As particles quickly move down the surface, tiny Brownian force perturbation can dramatically impact the trajectory path down to the bottom. For example, in the torus, the particles might fall down either in the front or the back surface even they start with the same initial position [Fig. 4(e)]. Likewise, particles on top of a ring of a knot can move down along different tube surface and reach different parts at the bottom [Fig. 4(f)]. Compared with particle transport in Euclidean space, the higher sensitivity to Brownian motion can be possibly employed to engineering applications like switches and particle sorting devices.

**Multiple particle dynamics on complex surfaces**

Dynamics of multiple particles on curved surfaces have generate significant excitement for both theoretical exploration[14, 21] and its potential to aid in engineering novel structures and devices[13, 15, 16]. As a final demonstration of our algorithm, we simulate colloidal particle assembly on curved surfaces under external force field [Fig. 5]. We simulate multiple colloidal particle with mild attraction levels (~$5k$T) on the surfaces and apply the external force field with $F_z = 10k$T/$a$ (e.g. gravity). In a flat 2D plane, crystallization of such weakly attractive colloidal particles require radial inward force to compress and concentrate the colloids, otherwise entropy will drive all particles to sample the whole space[41, 42]. In our example, the external forces, although acting in the z direction, also concentrate particles towards the bottom and induce crystallization. Fig. 5(a) and (b) shows the trajectories of multiple particles, starting from a uniformly distributed initial configuration, moving toward the bottom and finally forming a solid crystalline structure [Fig. 5(c) and (d)].

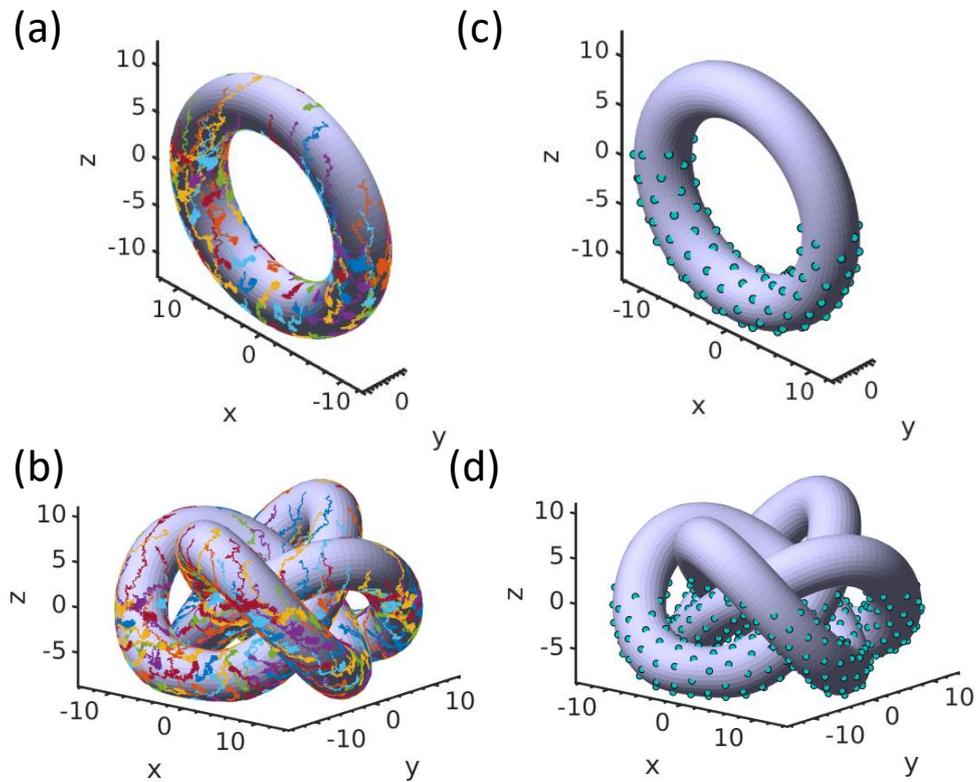

**Figure 5**. (a, b) Assembly trajectory and equilibrium configuration of 120 particles on a torus surface under external field in the negative z direction characterized by $F_z$ = -10kT/a. (c, d) Assembly trajectory and equilibrium configuration of 480 particles on a knot surface under the same external field as in (a).

At equilibrium, besides predominant hexagonal close-packed structures formed on the torus and the knot surface, we also identify topological defects including mainly heptagons (seven-fold coordinated particle) and pentagons (five-fold coordinated particle) [Fig. 6(a) and (b)]. Pentagons and heptagons facilitate the tiling of a curve surface (e.g., soccer balls), and they are commonly found in crystals on curved surfaces[6, 20]. We proceed to quantify the impact of curvature and topology on the crystalline packing structure by analyzing the pair correlation distribution $g(r)$ [Fig. 6(c–e)], whose value can be interpreted as the probability (up to a scaling factor) to find another particle at the given distance $r$ from the center of one particle. For comparison, we also compute $g(r)$ for an equilibrated finite-size circular crystalline lattice (300 particles) on a 2D plane obtained via electric field compression[41, 43, 44].

As particles predominately forming hexagonally close-packed (HCP) structure on surfaces, all $g(r)$ functions exhibit similar peak locations corresponding to the first four coordinate shells. There are several key observations. First, for lattices on a plane, $g(r)$ has sharper peaks and vanishing values between peaks, whereas on curved surface, the curvature makes particles pair distance to have larger variation and, as a result, $g(r)$ exhibits wider and shorter peaks. Also note that the first peak $g(r)$ for the torus surface is shaper than that of the knot surface, because the knot surface is more complex and has larger curvature variation than the torus surface (i.e., more twists and smaller tubes), which cause more defects and imperfect HCP structures [Fig. 6(b)]. Besides the peaks at the positions of the first four coordinate shells, we observe an additional small and wide peaks at the position of ~$5a$. This can be attributed to the surface geometry where particles are packing on the tube with a diameter of $5a$.

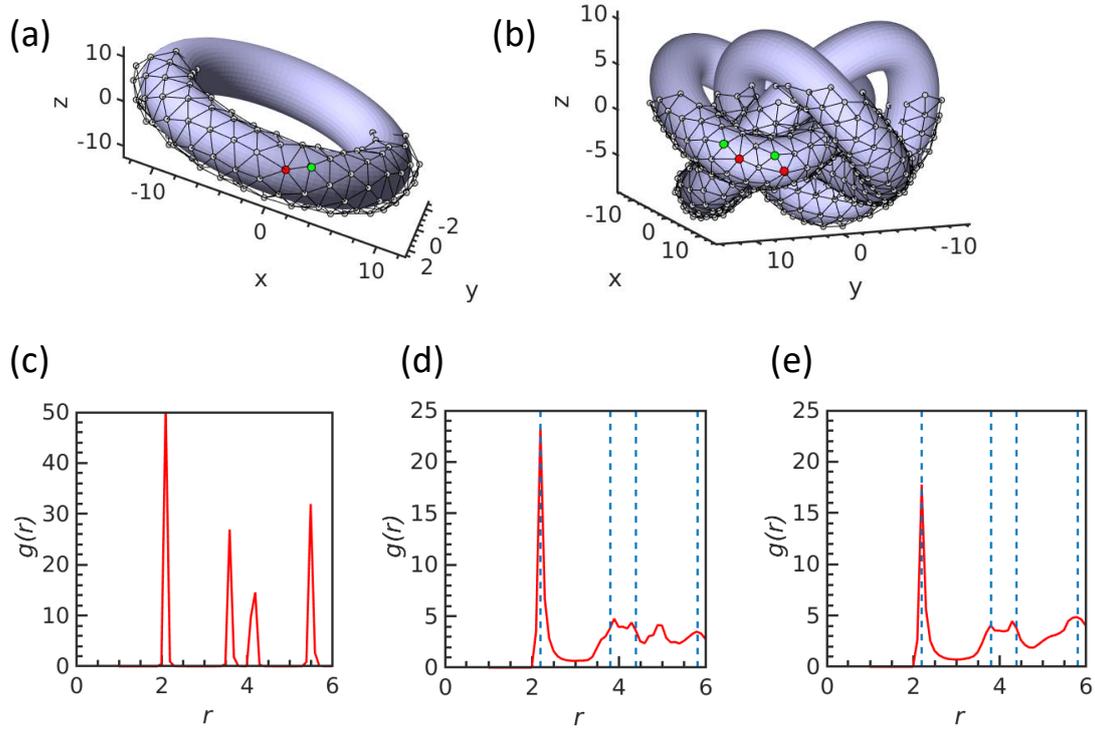

**Figure 6**. (a, b) Topological defects in the equilibrium structure of particle assembly on the torus (a) and the knot (b). Particles colored red are heptagons (seven-fold coordinated particle) and particles colored green are pentagons (five-fold coordinated particle). The neighbors for each particle is identified by Voronoi diagram. (c) Pair distribution function g(r) for an equilibrium finite-sized circular-shape hexagon close-packed lattice on a flat plane. (b, c) Pair distribution function g(r) for the equilibrium structure in (a) and (b). Dashed lines are the positions of peaks in g(r) for a 2D hexagonally-close-packed crystal in (c). The positions of the dashed lines are shifted to align with the first peaks in (d) and (e).

## Conclusion and Outlook

In this work, we have developed a BD simulation algorithm for particles moving on general curved surfaces. By approximating surfaces via triangle meshes and performing constrained Brownian simulation on the meshes, we can simulate Brownian particles dynamics on surfaces with complex geometry and topology. We have verified the simulation algorithm by comparing simulation results and theoretical predictions on a 2D flat surface and a spherical surface. We demonstrate the application of our algorithms in simulating the single particle diffusion and transport and crystallization of collective particles on a torus and a knot surface. Our results illustrate the impact of curvature and topology on particle dynamics and equilibrium structures.

Since the methodology to create and process triangle mesh surfaces is well-established and applied to approximate curved surface in broad areas, our algorithms will find application in a wide range of areas. For example, besides the study of point defects arises from curved geometry[3], our tools can also be applied to study grain boundaries[41, 44, 45], glasses[46-49], jamming[50, 51] in colloidal systems on curved surfaces. It can also be used as a flexible tool to simulate proteins on membrane with complex geometry[7, 8] and viral assembly [52]. The algorithm can also be applied to active systems to study the impact of geometry and topology on the non-equilibrium dynamics, including lining, jamming, clustering, and oscillation[53-56]. Although we focus on the impact of geometry and topology and ignore hydrodynamics, the simulation algorithm can be extended to include interfacial hydrodynamics[28, 57, 58] by adjusting the diffusivity terms in Eq. (5).

## Acknowledgement

Support from National Natural Science Foundation of China (Grant No. 11672161) is acknowledged. We also want to thank Prof. Misha Kazhdan and Prof. Gregory Chirikjian at Johns Hopkins University for inspiring discussion.

## Conflict of Interest Disclosure

The authors declare no competing financial interest.